\RequirePackage[running]{lineno}

\documentclass[aps,prl,nofootinbib,twocolumn,showpacs,superscriptaddress,letterpaper,amsmath,amssymb]{revtex4}

\pdfoutput=1

\usepackage{graphicx}  
\usepackage{dcolumn}   
\usepackage{bm}        
\usepackage{amssymb}   
\usepackage{feynmf}
\usepackage{slashed}
\usepackage{multirow}
\def\gsim{\lower0.5ex\hbox{$\:\buildrel >\over\sim\:$}}
\def\lsim{\lower0.5ex\hbox{$\:\buildrel <\over\sim\:$}}

\newcommand{\be}{\begin{equation}}
\newcommand{\ee}{\end{equation}}
\newcommand{\bea}{\begin{eqnarray}}
\newcommand{\eea}{\end{eqnarray}}

\newcommand{\nbox}{{\,\lower0.9pt\vbox{\hrule \hbox{\vrule height 0.2 cm
\hskip 0.2 cm \vrule height 0.2 cm}\hrule}\,}}

\def\ie{{\it i.e.}}

\def\sub#1{_{\lower.25ex\hbox{$\scriptstyle#1$}}}

\def\to{\rightarrow}

\newskip\zatskip \zatskip=0pt plus0pt minus0pt
\def\matth{\mathsurround=0pt}
\def\lsim{\mathrel{\mathpalette\atversim<}}
\def\gsim{\mathrel{\mathpalette\atversim>}}
\def\sigv{\ifmmode \langle\sigma v\rangle\else $\langle\sigma v\rangle$\fi}
\newskip\zatskip \zatskip=0pt plus0pt minus0pt
\def\matth{\mathsurround=0pt}
\def\lsim{\mathrel{\mathpalette\atversim<}}
\def\gsim{\mathrel{\mathpalette\atversim>}}
\def\atversim#1#2{\lower0.7ex\vbox{\baselineskip\zatskip\lineskip\zatskip
  \lineskiplimit
  0pt\ialign{$\matth#1\hfil##\hfil$\crcr#2\crcr\sim\crcr}}}

\def\missET {{\not\!\! E_T}}

\begin{document}

\thispagestyle{empty}
\vspace*{-3.5cm}

\vspace{0.5in}

\title{Collider Bounds on Indirect Dark Matter Searches: the $WW$
  final state}

\begin{center}
\begin{abstract}
We describe an effective theory of interaction between pairs of dark
matter particles (denoted $\chi$) and pairs of $W$ bosons. Such an interaction could
accomodate $\chi\bar{\chi}\rightarrow WW$ processes, which are a major
focus of indirect dark matter experiments, as well as $pp \rightarrow W\rightarrow W\chi\bar{\chi}$ processes, which would
predict excesses at the LHC in the $W+\missET$ final-state.  We reinterpret
an ATLAS $W+\missET$ analysis in the hadronic mode and translate the
bounds to the space of indirect detection signals. We also reinterpret the $W+\missET$ analysis in terms of graviton theory through the processes $W\rightarrow WG$ and $Z\rightarrow ZG$ in which $G$ is invisible. Finally, the final state is interpreted in terms of a $W'$ model where $W'\rightarrow WZ$, where $W$ decays hadronically and $Z$ decays to neutrinos.

\end{abstract}
\end{center}

\author{Nicolas Lopez}
\affiliation{Massachussets Institute of Technology, Cambridge, MA}
\author{Linda M. Carpenter}
\affiliation{The Ohio State University, Columbus, OH}
\author{Randel Cotta}
\affiliation{Department of Physics and Astronomy, University of
  California, Irvine, CA 92697}
\author{Meghan Frate}
\affiliation{Department of Physics and Astronomy, University of
  California, Irvine, CA 92697}
\author{Ning Zhou}
\affiliation{Department of Physics and Astronomy, University of
  California, Irvine, CA 92697}
\author{Daniel Whiteson}
\affiliation{Department of Physics and Astronomy, University of
  California, Irvine, CA 92697}

\maketitle


\subsection{Introduction}

It is well-established that dark matter makes up an significant fraction of the matter and energy density of the Universe~\cite{dmReview,planck}, but its particle nature and the form of its non-gravitational interactions remain important mysteries.

Dedicated experiments search for interactions between dark matter particles and quiet nuclei (called {\it direct detecton}~\cite{lux,xenon}), or for dark matter annihilation in space leading to visible particles ({\it indirect detection}~\cite{fermi}).  In addition, experiments at high-energy particle colliders play a complementary role, often with the greatest sensitivity for low-mass dark matter~\cite{atlasjet,cmsjet, atlasphoton,cmsphoton}. In order to analyze the collider data in the same framework as indirect- and direct-detection experiments, it is convenient to encapsulate our lack of knowledge of the form of the interaction between dark matter and the standard model as an effective field theory (EFT), in which the dark matter is fairly light and mediators are heavy enough to be integrated into four-point effective vertices~\cite{Goodman:2010ku,Beltran:2010ww,Fox:2011pm} between quarks and dark matter particles.

At particle colliders, this interaction produces invisible pairs of dark matter particles ($pp\rightarrow \chi\bar{\chi}$), and so relies on initial-state radiation of a visible object (jet, photon, $Z$ boson, etc) in order to leave a visible detector signature (jet+$\missET$, $\gamma+\missET$, $Z+\missET$, etc).  In these cases, initial-state radiation of a photon or $Z$ boson is not as sensitive as radiation of a gluon~\cite{dmcombo} for theories in which dark matter interacts with quarks and gluons, but these channels have unique power to probe interactions between DM and photons or $Z$ bosons, leading to effective vertices with two gauge bosons and two DM particles~\cite{monoz,Bell:2012rg, Nelson:2013pqa,Cotta:2012nj,Rajaraman:2012fu}.

\begin{figure}
\includegraphics[width=1.5in]{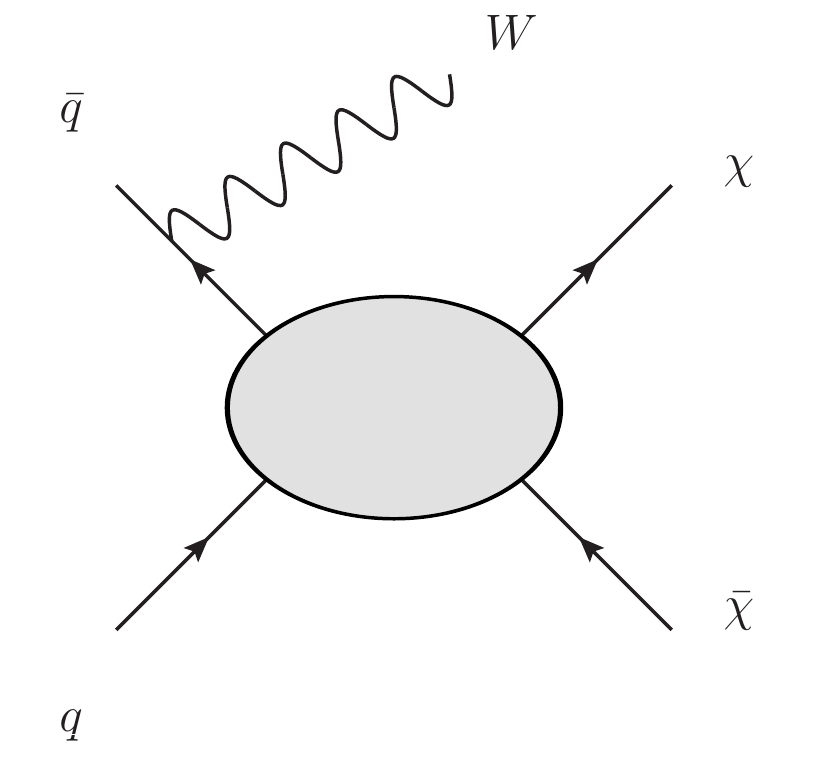}\\
\includegraphics[width=1.65in]{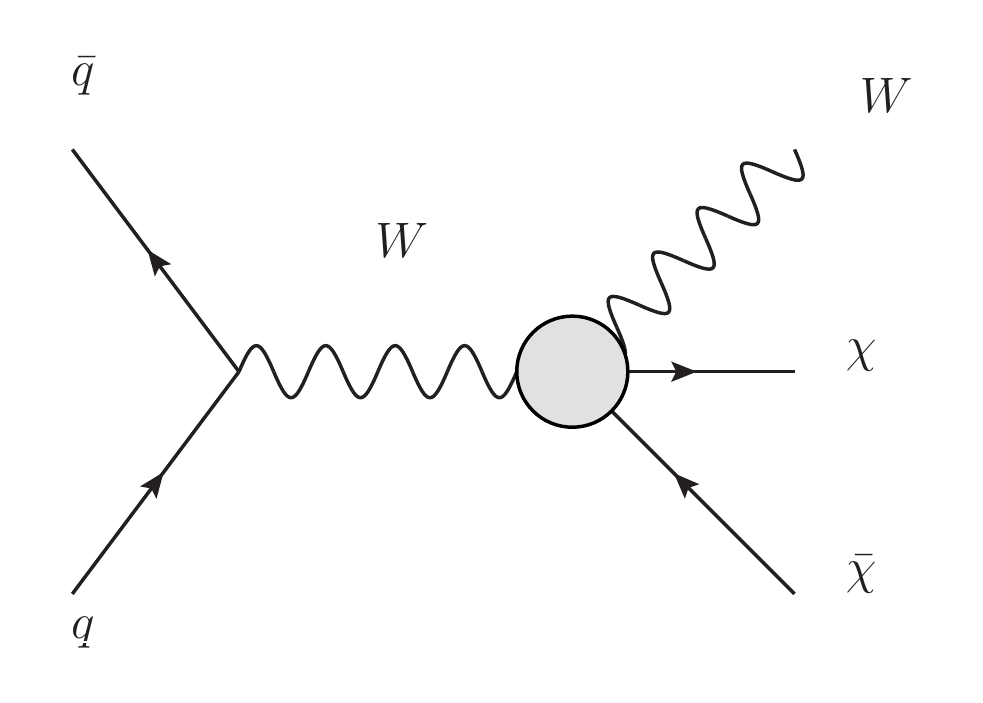}
\caption{\label{diag}Representative diagrams for production of dark matter pairs ($\chi \bar{\chi}$) associated with
a $W$ boson in theories where dark matter interacts with quarks (top) or directly with weak boson pairs (bottom). The latter
are those that we consider in this work.}
\end{figure}

In this paper, we extend this line of thought to the $W+\missET$
final state. We reinterpret the recent ATLAS analysis~\cite{atlaswhad} which sets limits on
theories of quark-DM effective interactions (see the top of
Fig.~\ref{diag}) in terms of theories of
$W$-DM effective interactions (see the bottom of
Fig.~\ref{diag}), working in an effective theory framework with a very
simple parameter space.

This class of interactions is of particular interest as collider
production of $W+\missET$ via $W\rightarrow W\chi\bar{\chi}$
is tied directly to the signal rates of indirect dark matter
signals via $\chi\bar{\chi}\rightarrow WW$, allowing the comparison
of LHC and indirect experiments in the parameter space of our
new effective theory.

In addition, we point out that the collider $W+\missET$ signature has broad sensitivity to other models which produce invisible particles. We demonstrate how this extends to graviton production in ADD~\cite{ArkaniHamed:1998rs} models, as well as exotic $W'$ bosons~\cite{Altarelli:1989ff}.

\subsection{Dark Matter Models}

In deriving bounds on effective theories of dark matter, we consider
effective operators through which pairs of neutral stable
particles (the DM) may couple to $W$ bosons. We consider the most relevant such
operators for both scalar and fermionic DM particles, denoted $\chi$.

Taking all operators of effective dimension 6 and 7, one finds that the DM couples
to the square of the field strength tensors of the $SU(2)$ gauge group through a
variety of Lorentz structures. The most general dimension 6 operators involving
scalar dark matter particles are:
 \bea
\label{eq:SMSinglet}
\mathcal{L}_{B1+B2}= \frac{1}{\Lambda_{B1}^2} ~\bar{\phi} \phi ~
F_1^{\mu \nu} F^1_{\mu \nu} \nonumber + \frac{1}{\Lambda_{B2}^2} ~\bar{\phi} \phi ~ F_2^{\mu \nu} F^2_{\mu \nu} \nonumber
\eea
and
\bea
\mathcal{L}_{B3+B4}= \frac{1}{\Lambda_{B3}^2} ~\bar{\phi} \phi ~   F_1^{\mu \nu} \tilde{F^1_{\mu \nu}} + \frac{1
}{\Lambda_{B4}^2} ~\bar{\phi} \phi ~   F_2^{\mu \nu} \tilde{F^2_{\mu \nu}}
\eea
\noindent
where $F_i$, $i=1,2$ are the Standard Model M $U(1)$ and,
$SU(2)$ field strength tensors.  $\Lambda_{Bi}$ and $\Lambda_{Ci}$ below are the effective cut-off scales of the operators, following  the
notation of Ref.~\cite{Rajaraman:2012fu}. In the second
set of operators, the dual field strength tensor appears.

Generic operators in which fermionic DM can couple to $SU(2)$ bosons in
gauge-invariant fashion start at dimension 7. There are now more operators
to consider due to the Lorentz structure of the DM bilinear:
\bea
\mathcal{L}_{C1+C2}= \frac{1}{\Lambda_{C1}^3} ~\bar{\chi} \chi ~  F_1^{\mu \nu} F^1_{\mu \nu}+\nonumber \frac{1}{\Lambda_{C2}^3} ~\bar{\chi} \chi ~  F_2^{\mu \nu} F^2_{\mu \nu},
\eea
\bea
\mathcal{L}_{C3+C4}= \frac{1}{\Lambda_{C3}^3} ~\bar{\chi} \chi ~   F_1^{\mu \nu} \tilde{F^1_{\mu \nu}} + \frac{1}{\Lambda_{C4}^3} ~\bar{\chi} \chi ~   F_2^{\mu \nu} \tilde{F^2_{\mu \nu}}
\eea
as well as
\bea
\mathcal{L}_{C5+C6}= \frac{1}{\Lambda_{C5}^3} ~\bar{\chi} \gamma^5 \chi ~  F_1^{\mu \nu} F^1_{\mu \nu}+\nonumber \frac{1}{\Lambda_{C6}^3} ~\bar{\chi} \gamma^5 \chi ~  F_2^{\mu \nu} F^2_{\mu \nu},
\eea
\bea
\mathcal{L}_{C7+C8}\frac{1}{\Lambda_{C7}^3} ~\bar{\chi} \gamma^5 \chi ~   F_1^{\mu \nu} \tilde{F^1_{\mu \nu}} + \frac{1}{\Lambda_{C8}^3} ~\bar{\chi} \gamma^5 \chi ~   F_2^{\mu \nu} \tilde{F^2_{\mu \nu}},
\label{C56}
\eea

Importantly, however, the $SU(2)$ invariance of the dimension 6,7 operators
mentioned above implies precise relationships between operators connecting the
DM particles with various electroweak gauge bosons of the SM.

These couplings are presented using a modified notation which nicely demonstrates how DM couplings to gauge bosons are related by gauge invariance.
For effective operators involving $\mathcal{L}_{Cn + Cn+1}$ we define $\left(1/{\Lambda_{Cn}}^3\right)$ = $k_1/\Lambda^3$ and $\left(1/\Lambda_{Cn+1}^3\right)$ = $k_2/\Lambda^3$. Similarly, for any effective operators involving $\mathcal{L}_{Bn + Bn+1}$ we define $\left(1/\Lambda_{Bn^2}\right)$ = $k_1/\Lambda^2$ and $\left(1/\Lambda_{Bn+1}^2\right)$ = $k_2/\Lambda^2$.  The DM couplings to pairs of electro-weak bosons are thus given by:
\bea
g_{WW}&=&\frac{2k_2}{s_w^2 \Lambda^{2-3}} \\\nonumber
g_{ZZ} &=& \frac{1}{4 s_w^2 \Lambda^{2-3}} \left(\frac{k_1 s_w^2}{c_w^2}+\frac{k_2 c_w^2}{s_w^2} \right) \\\nonumber
g_{\gamma\gamma}&=&\frac{1}{4 c_w^2}\frac{k_1+k_2}{\Lambda^{2-3}} \\\nonumber
g_{Z\gamma} &=& \frac{1}{2 s_w c_w \Lambda^{2-3}} \left(\frac{k_2}{s_w^2}-\frac{k_1}{c_w^2} \right)
\label{eq:prefactors}
\eea
\noindent
where $s_w$ and $c_w$ are the sine and cosine of the weak mixing angle, respectively.
In all cases the overall operator coefficient coupling DM particle pairs
to pairs of $W$ bosons depends only on a single parameter $k_2/\Lambda^2$ (or
 $k_2/\Lambda^3$ depending on the dimension of the operator). Assuming then
that a single operator structure dominates, the overall production
cross section of $ pp \rightarrow W \chi \chi$ will depend only on two
parameters, the mass of the dark matter particle and the overall coefficient
$k_2/\Lambda^{2,3}$.  A prediction of a specific number of $W+\missET$ events also implies correlated
predictions for numbers of events in other mono-boson channels as well.

\subsection{Heavy Boson Models}

In the case of gravitons, we consider the ADD model of extra dimensions\cite{ArkaniHamed:1998rs}, which is a proposed solution to the hierarchy problem in which there are  $\delta$ extra dimensions of submilimeter size through which gravity propagates, with all other Standard Model (SM) fields localized on a higher dimensional 3 space brane. In this model, the reduced Planck scale $M_D$, where gravity becomes strong, is far below the 4-D Plank scale $M_P\sim10^{18}$ GeV.  The reduced Plank scale is set by the number and radius of the extra dimensions. $M_P^2$ = $R^{\delta}M_D^{2+\delta}$ where $R$ is the radius of extra dimensions. This model contains a series of Kaluza-Klein (KK) excitation states of graviton with masses $m_n=n/R$. Although each KK graviton has gravitational coupling, suppressed by $1/M_P$, the almost continous spectrum of KK gravtions are summed over and result in a coupling to SM particles suppressed only by powers of $M_D$. 

The LHC is a useful tool for exploring extra-dimensional scenarios in which the reduced Plank scale is is of order several TeV.  Note that for such low values of $M_D$ the ADD scenario with 1 extra dimension is clearly ruled out since it would involve extremely large extra dimensions.  For larger values of $\delta$ however, ADD scenarios are possible.  

In this case, the graviton $G$ is long-lived and invisible to the detector, such that the processes $Z\rightarrow GZ$ and $W\rightarrow GW$ give $Z+\missET$ and $W+\missET$ final states.

The $W'$ is a theoretical charged heavy vector boson which can decay to $WZ$. If the $Z$ decays to neutrinos, it gives the final state of $W+\missET$. The production cross section depends on a coupling of the form $\frac{m_{W}^{2}}{m_{W'}^{2}} \times g_{W'WZ'}$, meaning the coupling  will be inversely proportional to the mass of the $W'$ boson squared.  The $W'$ may also decay leptonically, although it is not discussed here.

\subsection{Experimental Search}
The ATLAS experiment at the LHC has placed limits on  dark matter
production in the $W+\missET$
channel~\cite{atlaswhad}, where the dark matter fields couple to quark initial states
and  the $W$ boson has been emitted as initial state radiation. These limits were derived from $20.3$ fb$^{-1}$ of data
produced in $pp$ collisions at $\sqrt{s}=8$ TeV. The full
selection is as follows:

\begin{itemize}
\item 1 Cambridge-Aachen jet with $R=1.2$, $p_{\textrm{T}}>250$ GeV, $|\eta|<1.2$, $\sqrt{y}> 0.4$
\item $\missET>350$ GeV
\item $\le 1$ narrow jet with $p_{\textrm{T}}> 40$ GeV,$|\eta|<4.5$,
  $\Delta R($narrow jet, fat jet$)>0.9$
\item No electrons, muons, or photons with $p_{\textrm{T}}>$ 10 GeV and
  $|\eta|<2.47$, $|\eta|<2.5$, and $|\eta|<2.37$ respectively
\end{itemize}

The results are consistent with the Standard Model expectation, as shown in Table~\ref{tab:atlasw}.

\begin{table}[tbh]
\caption{Data and estimated   background yields in the two signal
  regions, from Ref.~\cite{atlaswhad}. Uncertainties include statistical and systematic contributions.}
\center
\begin{tabular}{l|c|c}
\hline\hline
Process & $\missET>350$~GeV & $\missET>500$~GeV \\\hline
$Z\rightarrow\nu\bar{\nu}$                                    & $402 ^{+39}_{-34}$ & $54 ^{+8}_{-10}$ \\
$W\rightarrow \ell^\pm\nu$, $Z\rightarrow \ell^\pm\ell^\mp$   & $210 ^{+20}_{-18}$ & $22 ^{+4}_{-5}$ \\
$WW,WZ,ZZ$                                                    & $57 ^{+11}_{-8}$ & $9.1^{+1.3}_{-1.1}$ \\
$t\bar{t}$, single $t$                                        & $39 ^{+10}_{-4}$ & $3.7^{+1.7}_{-1.3}$ \\ \hline
Total                                                         & $707 ^{+48}_{-38}$ & $89 ^{+9}_{-12}$\\\hline
Data & 705 & 89 \\
\hline  \hline
\end{tabular}
\label{tab:atlasw}
\end{table}

Using the CLs method~\cite{cls1,cls2}, the ATLAS measurement
places an upper limit at 95\% confidence level on the cross section
within this fiducial region to be 4.4 fb, with a typical
reconstruction efficiency of 63\%.

In order to reinterpret these results
in terms of interactions with electroweak bosons, we need only
calculate the efficiency of the fiducial region selection for the theory of interest.

\subsection{Dark Matter Fiducial Efficiency and Limits}
\label{colliders}
We generate simulated samples of events for each hypothetical signal using {\sc madgraph}5~\cite{madgraph}, with showering and hadronization by {\sc pythia}~\cite{pythia}. With the exception of the jet-veto requirement, the  fiducial efficiency  can be reliably estimated using parton-level information.

\begin{figure}
\includegraphics[width=0.45\linewidth]{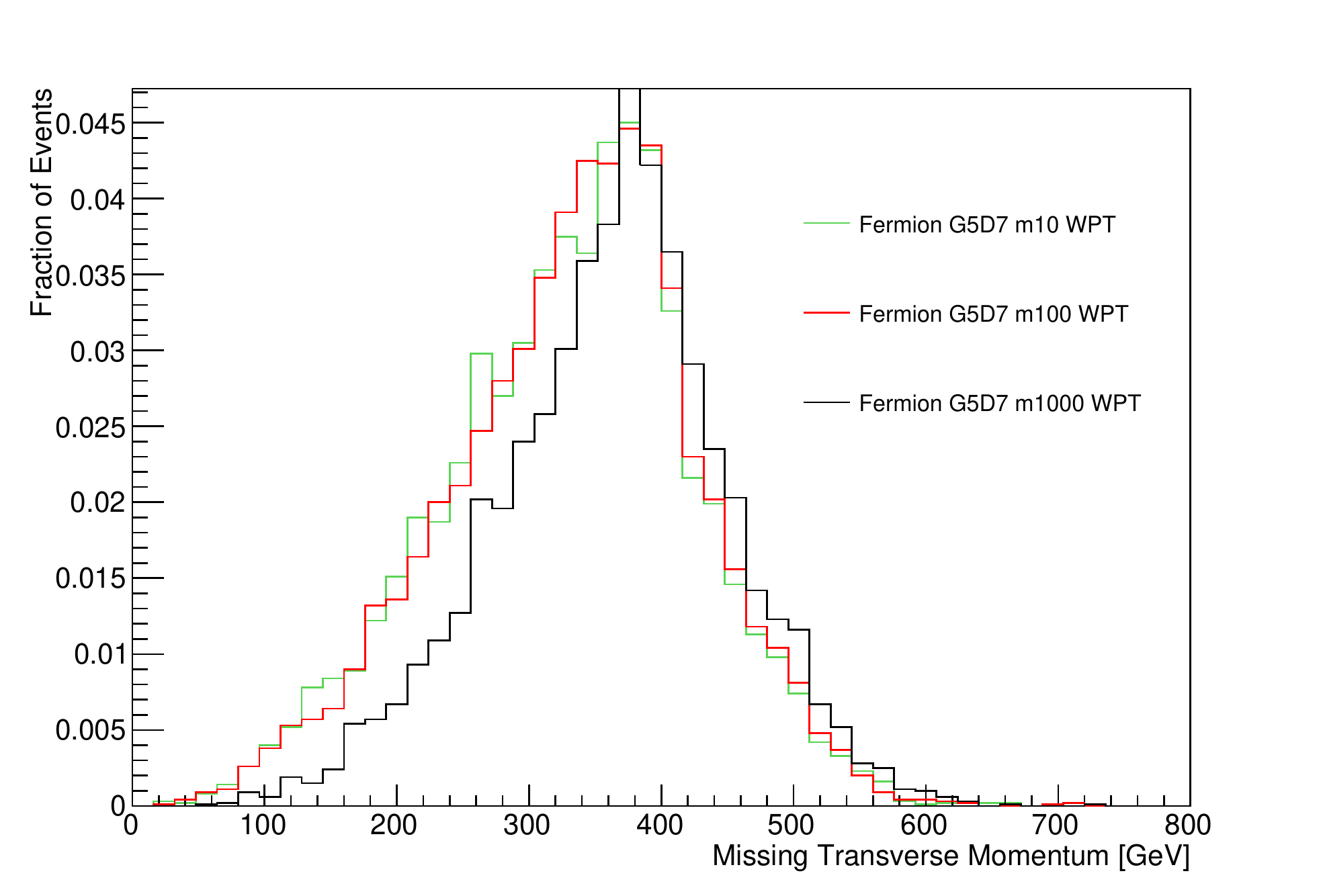}
\includegraphics[width=0.45\linewidth]{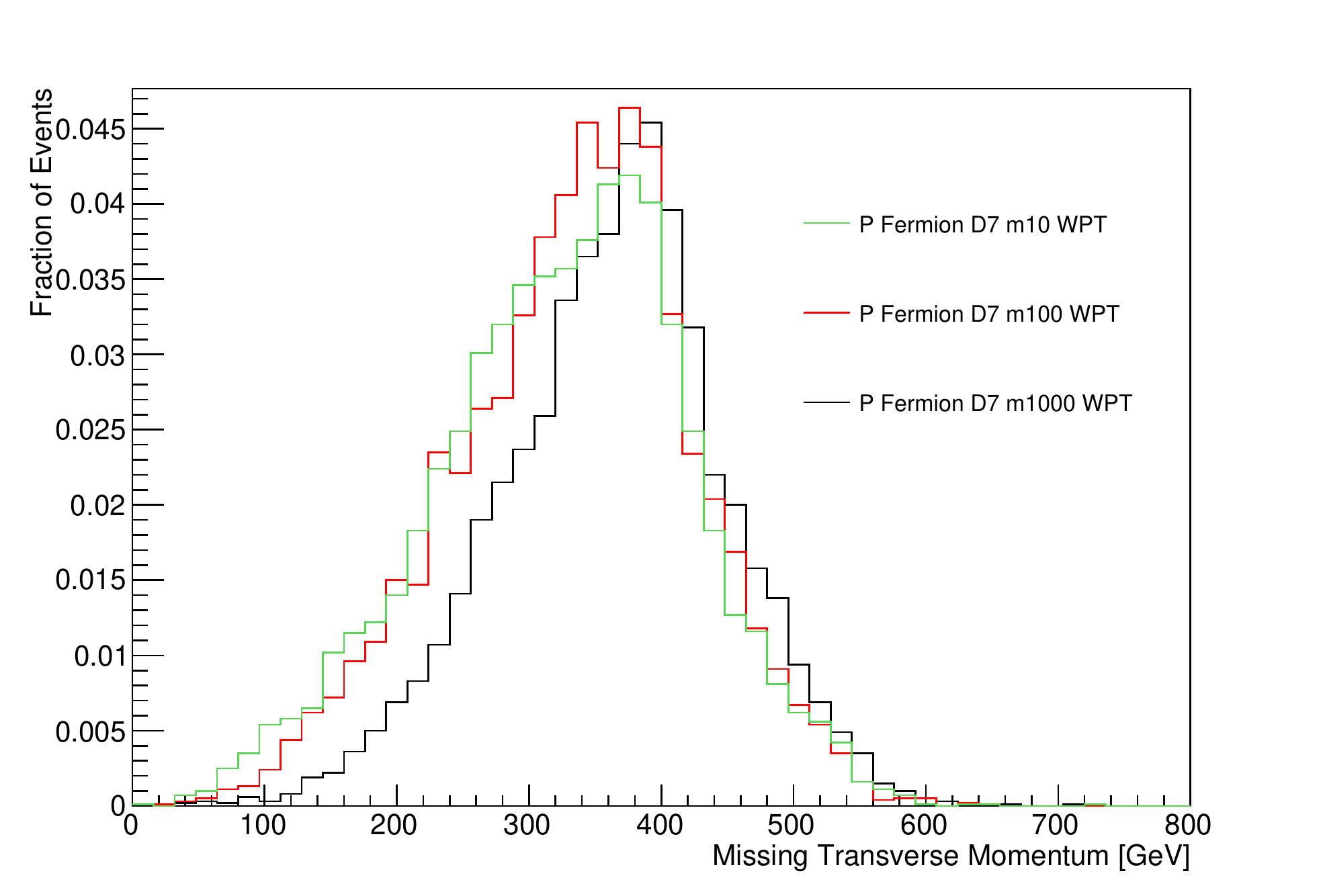}
\includegraphics[width=0.45\linewidth]{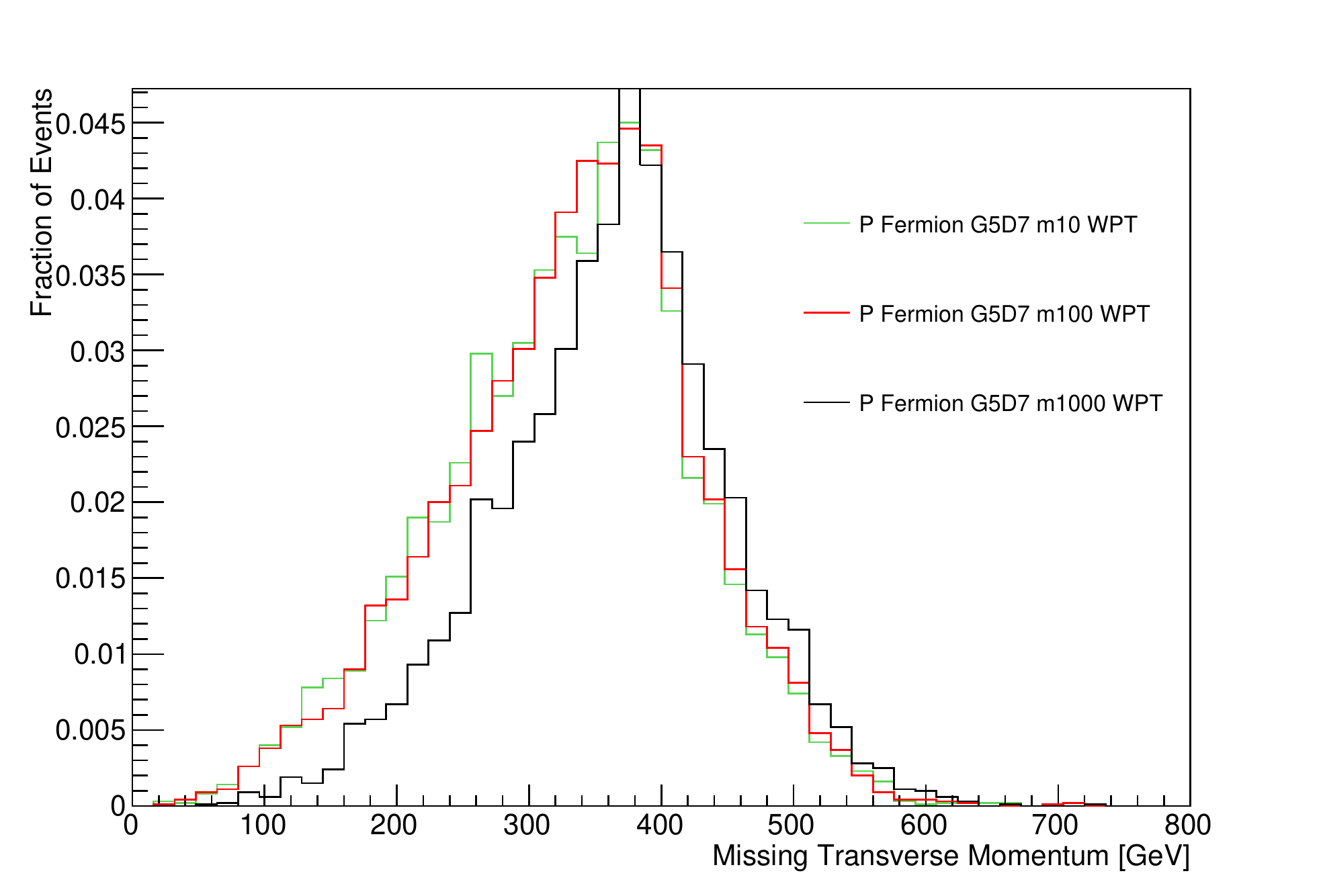}
\includegraphics[width=0.45\linewidth]{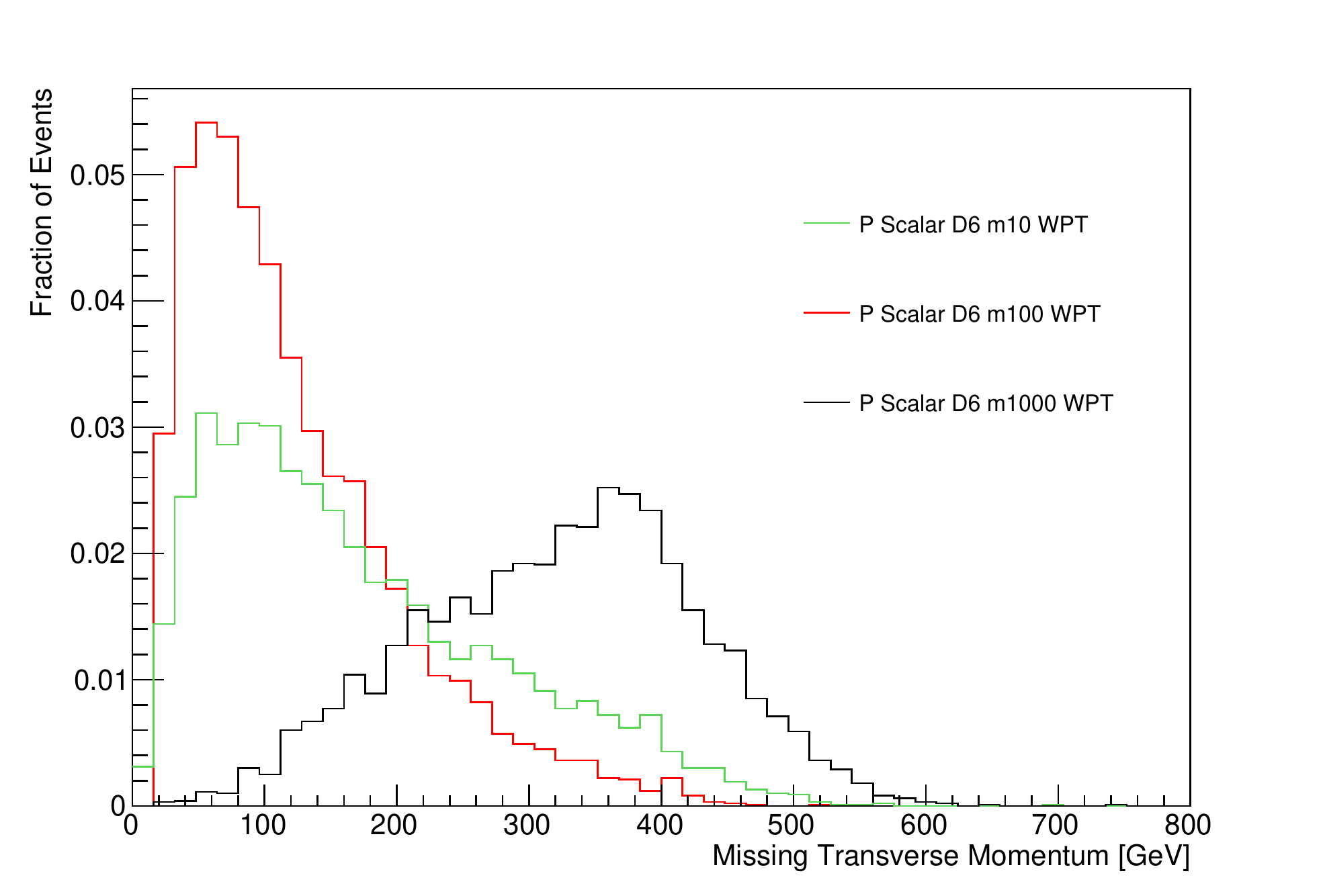}
\includegraphics[width=0.45\linewidth]{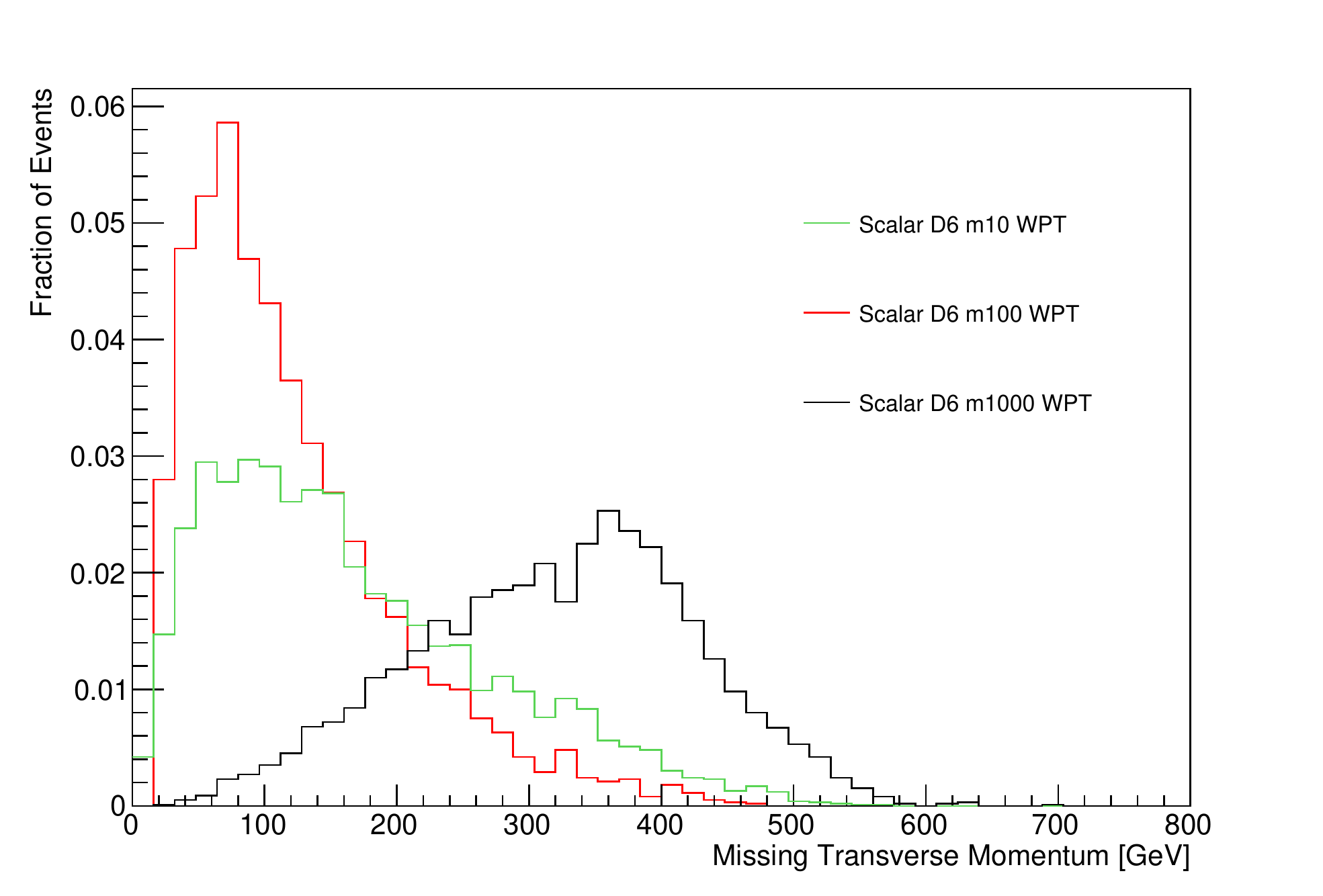}
\includegraphics[width=0.45\linewidth]{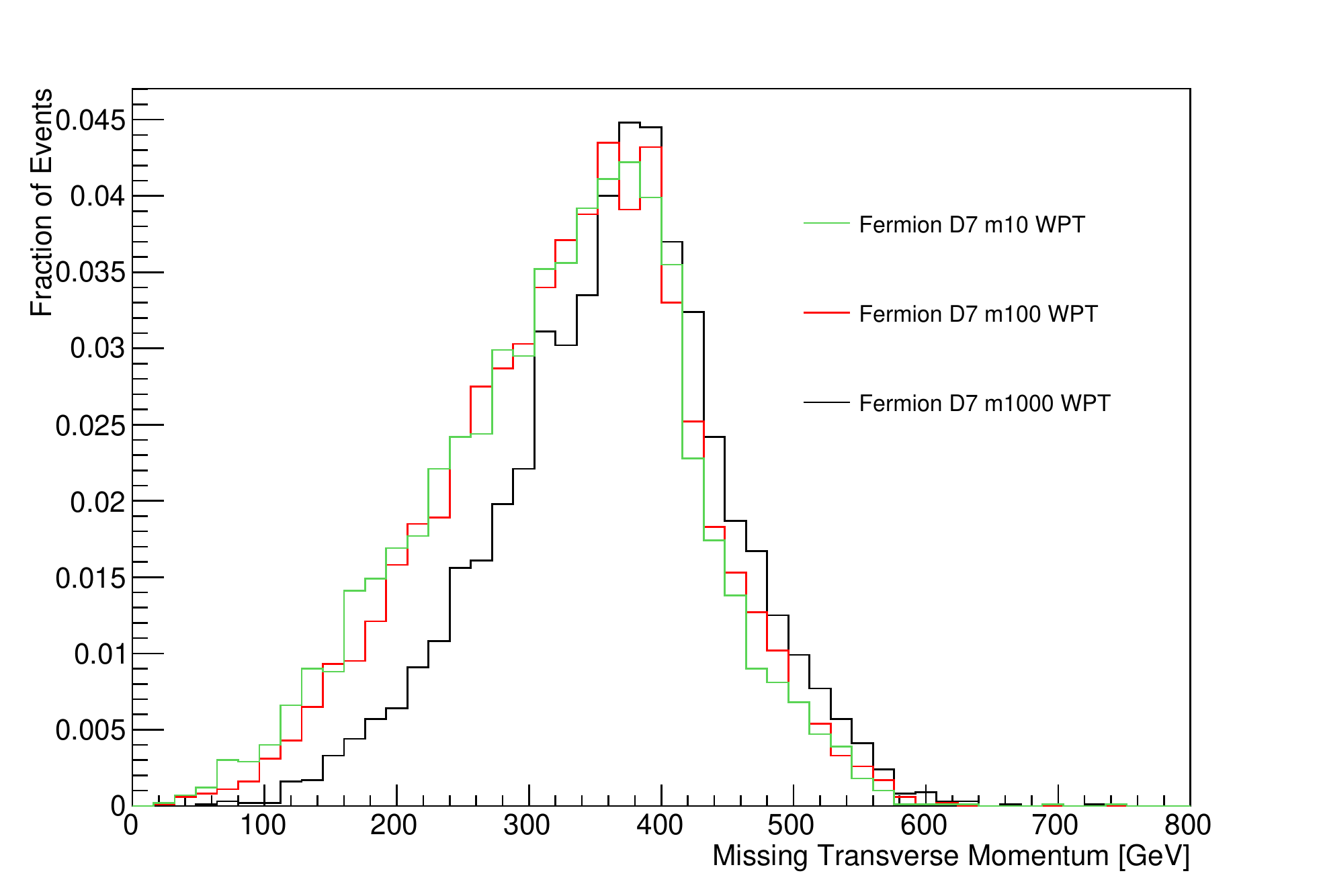}
\caption{ Distributions of $\missET$ in simulated $W\chi\bar{\chi}$
  events in $pp$ collisions at the  LHC for several choices of  $m_\chi$.}
\label{fig:met}
\end{figure}

In each case, the critical kinematic quantity which determines the efficiency is the missing transverse
momentum. Figure~\ref{fig:met} shows distributions of $\missET$ in simulated $W\chi\bar{\chi}$ events for our six effective field theories.

\begin{figure}
\includegraphics[width=0.9\linewidth]{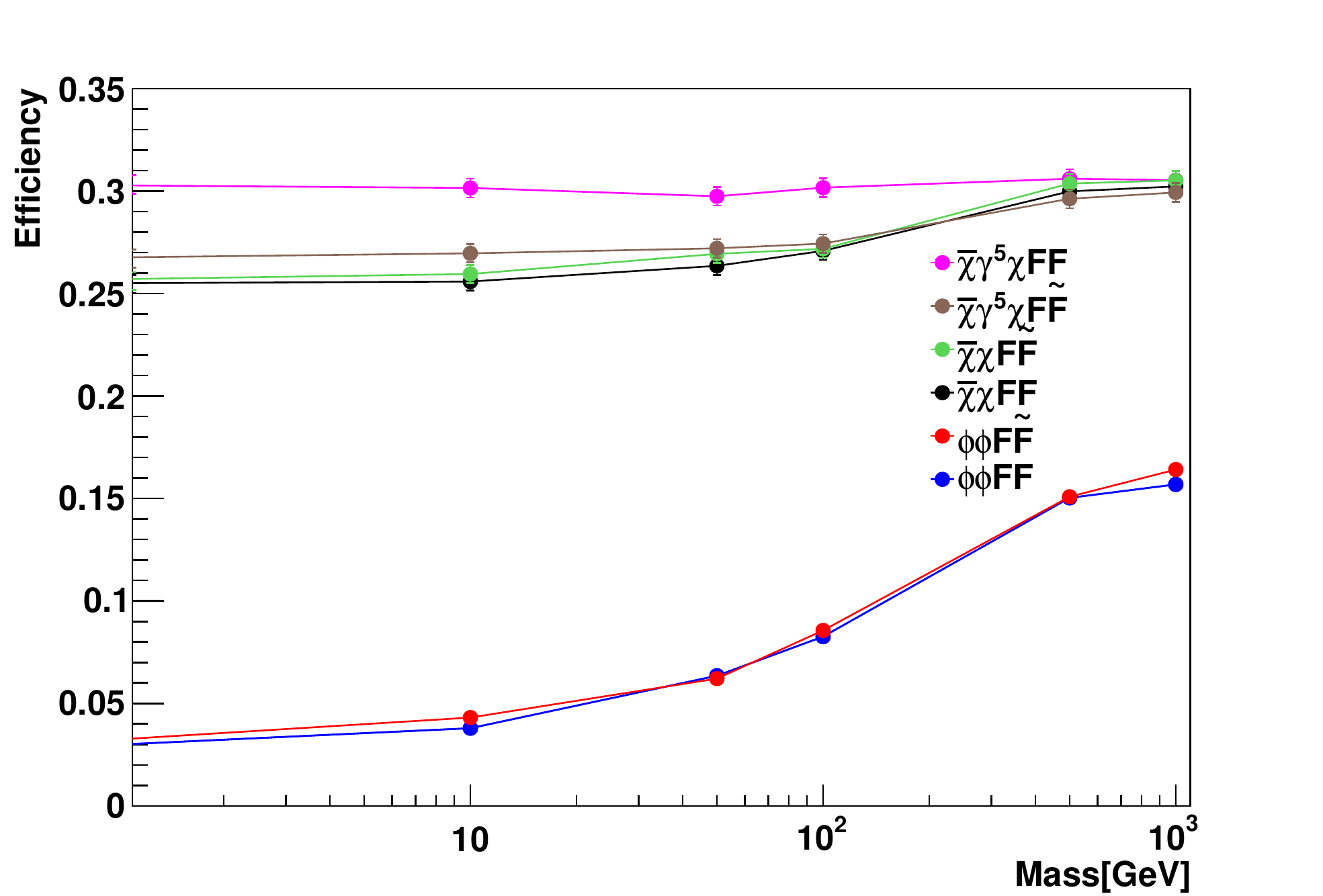}
\includegraphics[width=0.9\linewidth]{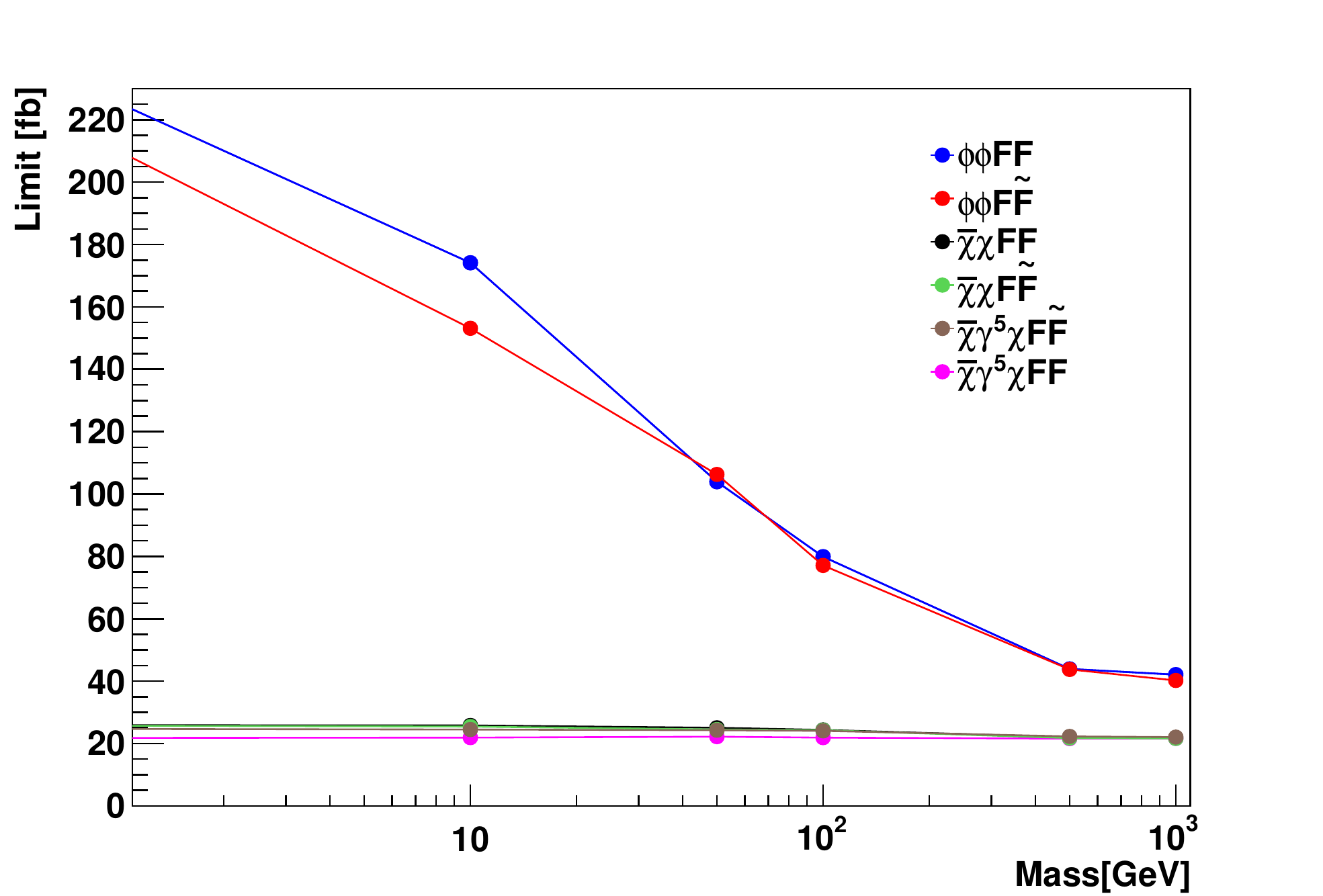}
\caption{ Fiducial efficiency and limits on $\sigma(pp\rightarrow W\chi\bar{\chi})$ for several
  values of $m_\chi$ for each of the EFTs describing interactions between DM and $W$ bosons.}
\label{fig:lim_mchi}
\end{figure}

\begin{figure}
\includegraphics[width=0.9\linewidth]{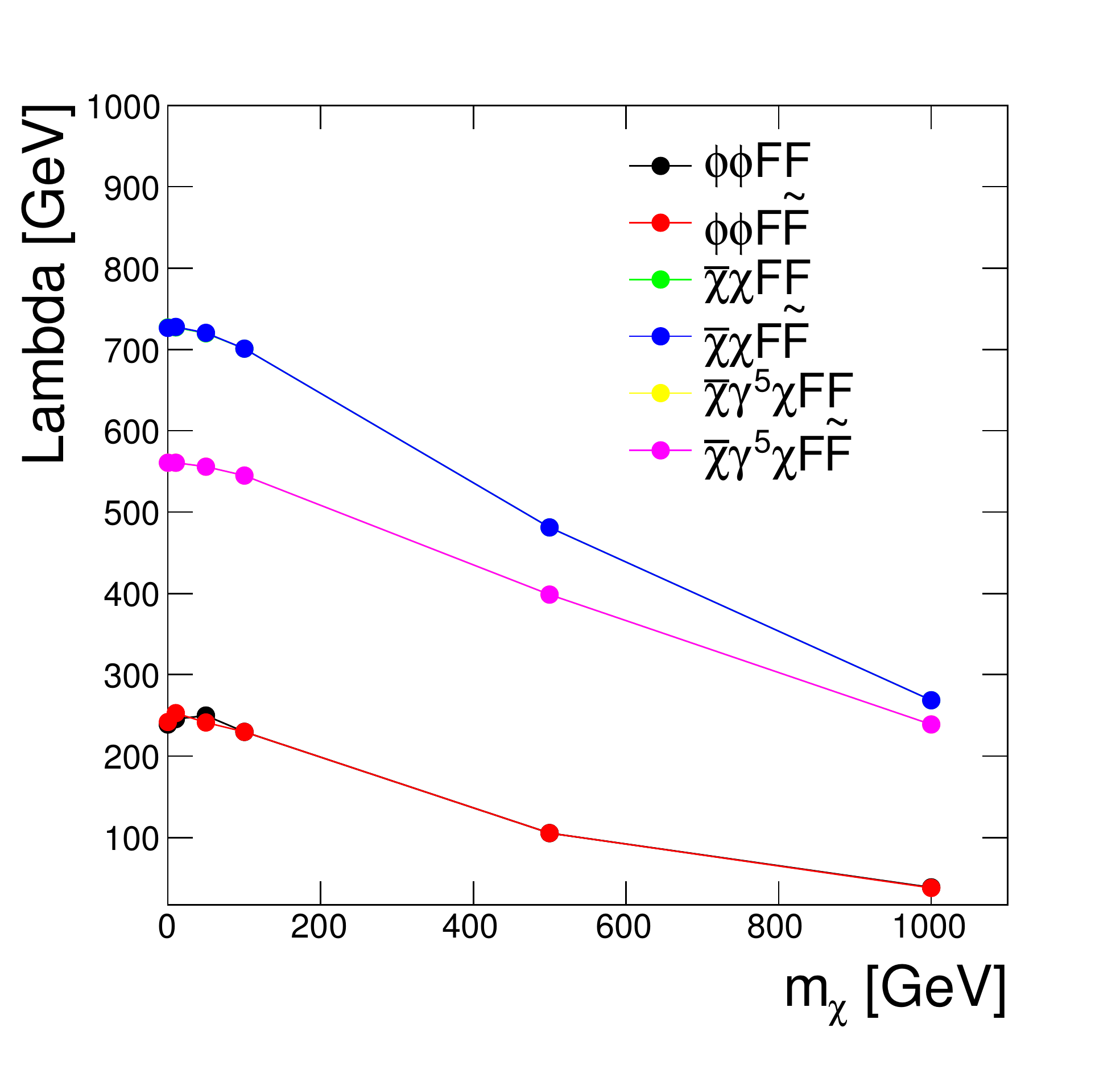}
\caption{ Limits on $\Lambda$ as a function of $m_\chi$.}
\label{fig:lim_lambda2}
\end{figure}

The fiducial efficiencies measured in these simulated samples allow us to calculate limits on the cross section, as shown   in Fig.~\ref{fig:lim_mchi} as a function of $m_\chi$. As the theoretical cross section depends on the suppression scale $\Lambda$, limits on the cross section can be translated into limits on $\Lambda$, see Fig ~\ref{fig:lim_lambda2}.

We observe (Fig.~\ref{fig:lim_mchi}) that the limits on light fermionic $\chi$ are much tighter than those
on light scalar $\phi$ DM, a feature that is obviously due to the differences in
$\missET$ spectra (Fig.~\ref{fig:met}). It is not hard to understand these differences as, in the limit of
massless $\chi$, the fact that the fermionic operators are dimension-7 and the scalar operators are
dimension-6 means that the cross-sections in the fermionic case must scale with a higher power of
the momenta involved\footnote{Terms that don't scale with momenta are much smaller, $\mathcal{O}(m^2_{\chi})/s$, $\ie$,
they are ``helicity suppressed.''}, and hence the $W$-boson $p_T$. The resulting cross-section is relatively suppressed as $p_T\to 0$
and is enhanced compared to the scalar case in the large $p_T$ tail.

\subsection{Indirect Signatures from our DM Operators}

In addition to collider production of $W+\missET$, our operators also mediate DM annihilation into the $WW$, $ZZ$, $\gamma Z$ and $\gamma\gamma$
final states. In this section, we convert results from indirect searches into limits on parameters of our EFTs.

The decay products, radiation and hadronization that follows production of the
final state bosons generically yields spectral continua of $\gamma$-rays, antiprotons,
positrons and neutrinos,
all of which are long known discovery channels in indirect searches for DM. The latter two processes
mediate direct annihilation to $\gamma$-rays, giving a highly distinctive monochromatic signal.

In our operator framework, bounds on the scale $\Lambda$ coming from indirect detection experiments
can be directly compared to the collider bounds derived above, provided some important
cosmological assumptions. The main assumption here is that our DM makes up
\emph{all} of the observed DM density in the universe. This is implicitly assumed in all of the bounds
derived below. A useful comparison for this assumption is to keep track of where the DM relic density
that \emph{would be} obtained from a thermal freezeout history, (\emph{assuming} that our operator is
the dominant annihilation process) matches, exceeds or undershoots the total relic density of the
universe. We provide contours of this thermal relic density calculation below so that we have an idea
of where we would need additional structure beyond our model in order to increase or decrease the
relic density to match our assumption that our DM makes up all of DM.

\begin{figure}
\includegraphics[width=3.2in]{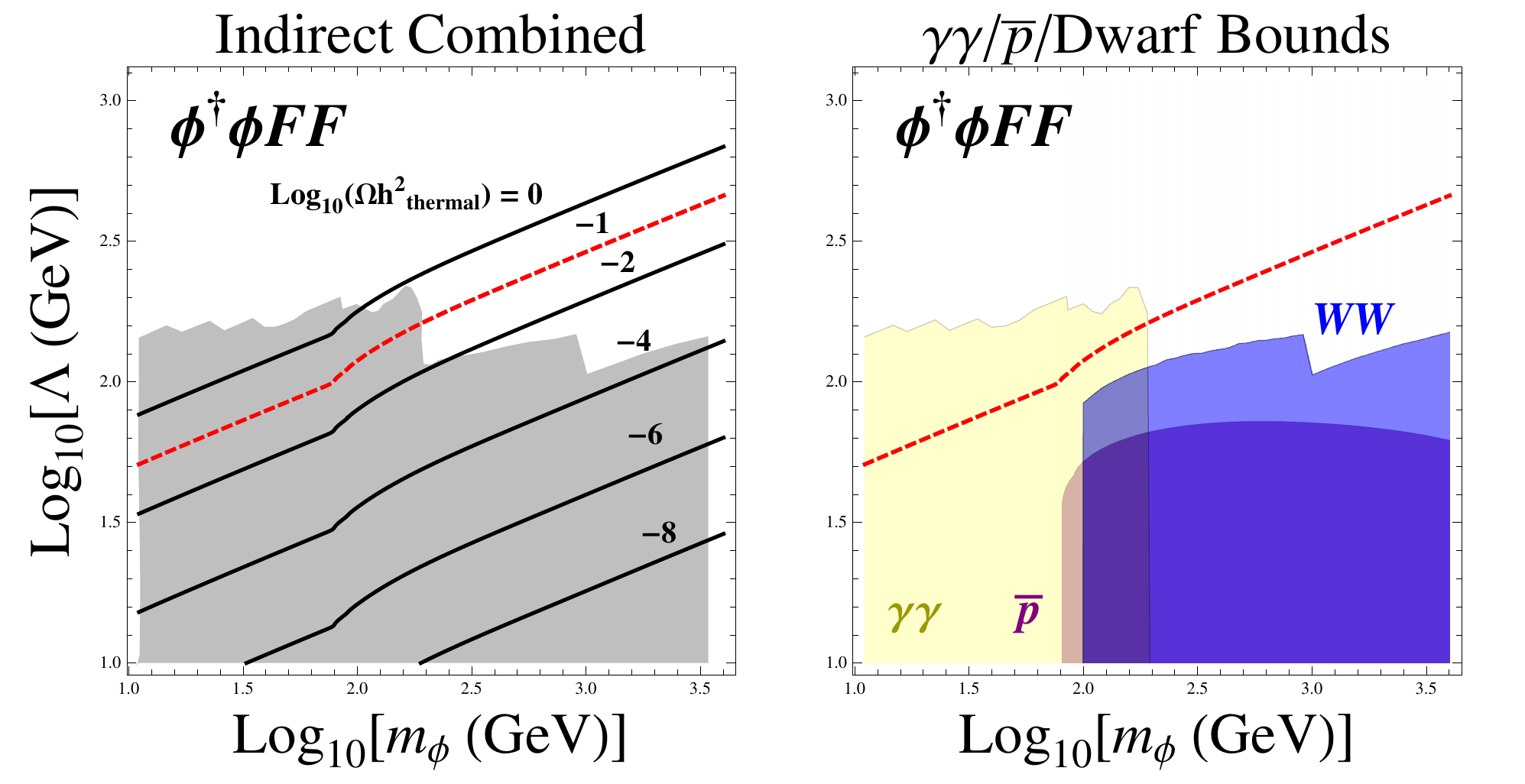}
\caption{
Bounds on the the $B_{1,2}$ operators from indirect detection searches for annihilation. Here $k_1=k_2$
so there is no bound from the monochromatic $\gamma Z$ channel. Shown are the constraints on the $\gamma \gamma$
channel from the LAT line search, on the $WW$ channel from dwarf bounds on continuum $\gamma$'s and from the
PAMELA $\bar{p}$ data. The black and red-dashed curves indicate the relic density that would be obtained from a
thermal calculation assuming this operator is the dominant process in freezeout. The picture is qualitatively similar
for the other operators so similar figures are omitted.
}
\label{indS6a}
\end{figure}

We combine the results of several experiments to derive our combined indirect bounds.
The analysis closely follows that of \cite{Cotta:2012nj}. The most robust bounds on continuum
$\gamma$-rays come from observations of dwarf galaxies. These are extremely
low background searches with high-quality estimates for the DM abundance and
distributions\footnote{we use
  limits that assume an NFW profile for the dwarf DM distribution, but these kinds of searches
(being sensitive to the integral of DM density squared over the entire distribution) are relatively
insensitive to this assumption.}. We also show bounds coming from the PAMELA antiproton
data \cite{Adriani:2010rc} (using the GALPROP\cite{Strong:1998pw}\cite{Moskalenko:2001ya}\cite{GALPROP}
propagation model galdef\_50p\_599278), although these bounds are relatively weak
in comparison
to the continuum $\gamma$-ray bounds. We use the bounds on the $WW/ZZ$ annihilation channels derived from
the \emph{Fermi}-LAT \cite{Ackermann:2011wa} and VERITAS \cite{Aliu:2012ga} data, as the two complement
each other in $m_{DM}$. Bounds on the monochromatic channels are those from the \emph{Fermi}-LAT work
\cite{Ackermann:2012qk} and assume an NFW profile for the galactic DM distribution (this is not as
robust an assumption as in the dwarf limits). The resulting bounds from indirect detection are shown
for the $B_{1,2}$ operators in Fig.~\ref{indS6a}.

We do not consider continuum $\gamma$-ray bounds from the galactic center (as they are exquisitely
sensitive to the DM distribution) or bounds from neutrino telescopes (bounds from the galactic
center are not competitive, while bounds from the solar DM search must rely on additional assumptions
about how our DM scatters on SM particles). There are no bounds from the $\gamma Z$ channel in our
plots as this channel is turned off for the custodially-symmetric combination $k_1=k_2$. Away from this
special case the \emph{Fermi}-LAT $\gamma Z$ bound would be a non-trivial constraint, however, we simply
note here that the reach would be much the same (and overlapping in $m_{DM}$) as the $\gamma\gamma$ bound.

\begin{figure}
\includegraphics[width=3.2in]{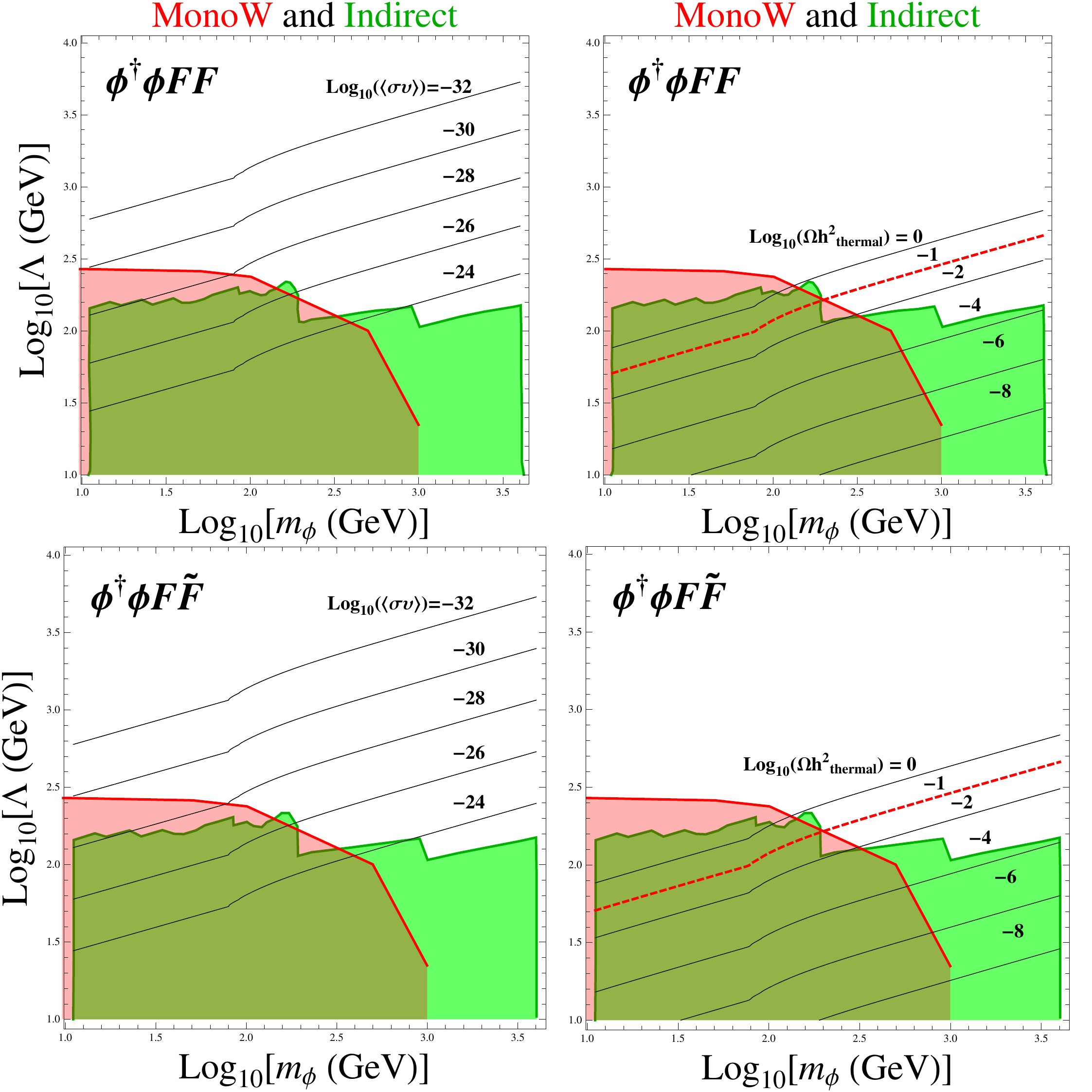}
\caption{
Comparison of constraints on the $B_{1,2}$ and $B_{3,4}$ operators coming from the collider $W+\missET$ search (red) and indirect searches (green).
Curves describing current annihilation cross-sections and thermal relic density are shown in the left and right
panels, respectively.
}
\label{comboS6ab}
\end{figure}

\begin{figure}
\includegraphics[width=3.2in]{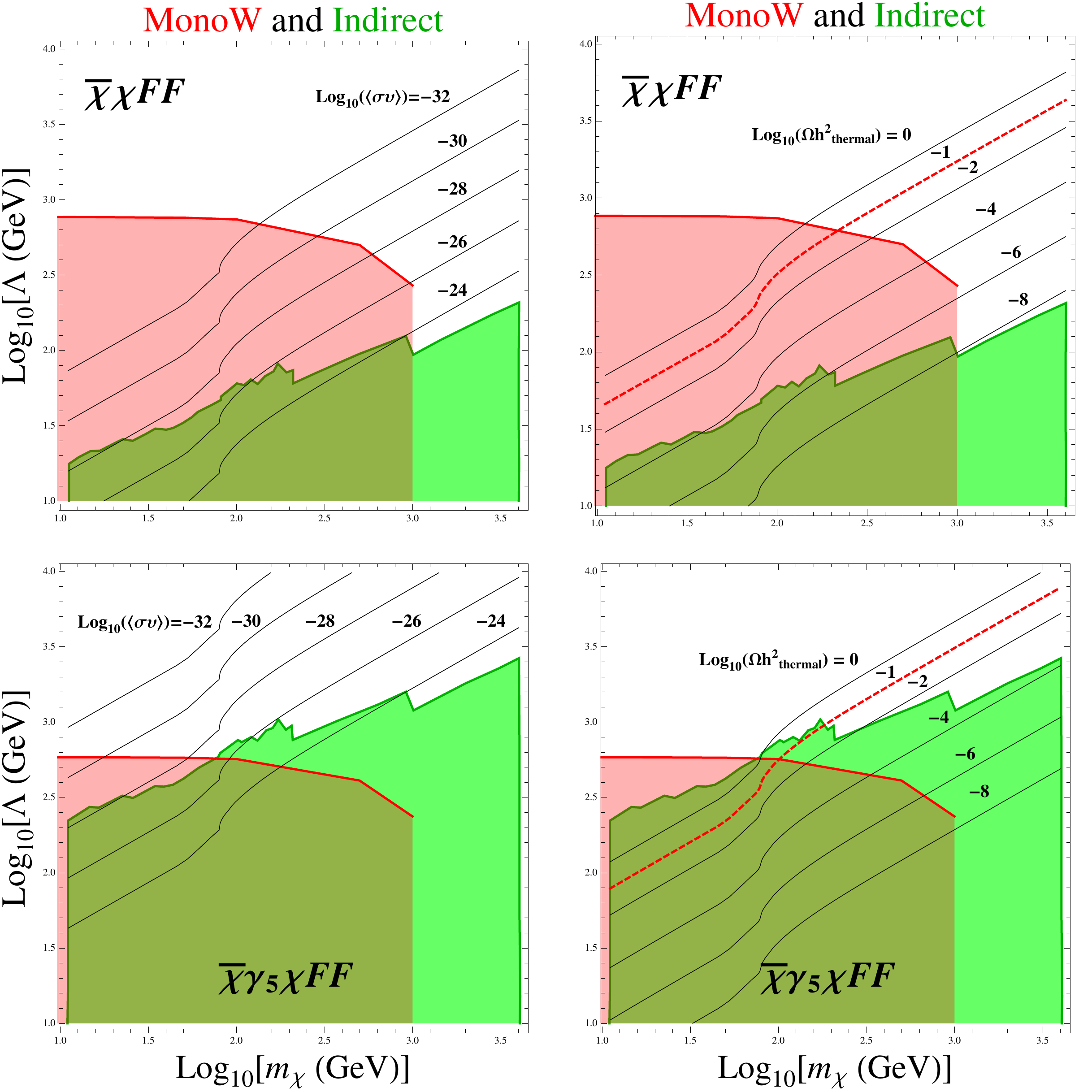}
\caption{
As in Fig.~\ref{comboS6ab} for the $C_{1,2}$ and $C_{5,6}$ operators.
}
\label{comboD7ab}
\end{figure}

\begin{figure}
\includegraphics[width=3.2in]{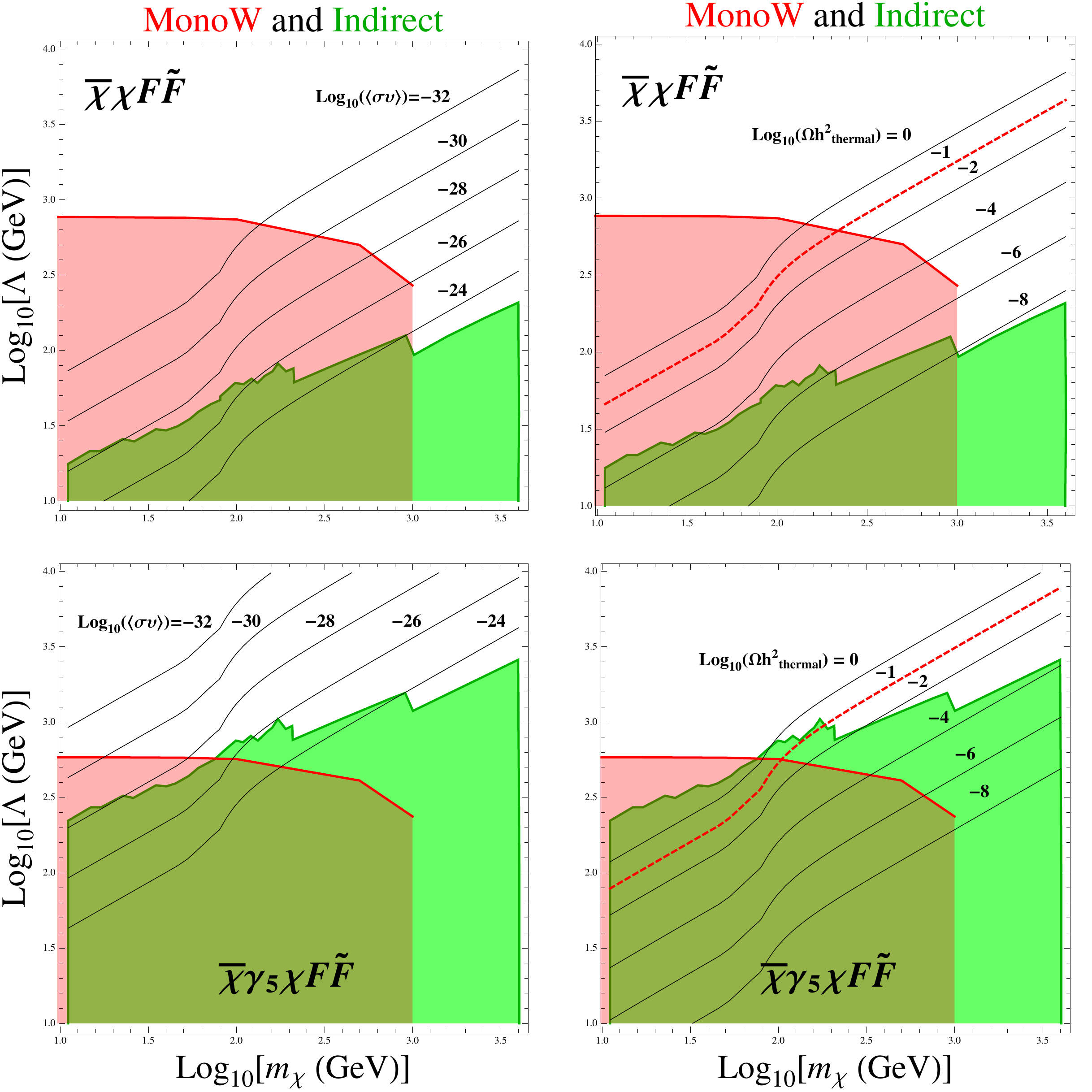}
\caption{
As in Fig.~\ref{comboS6ab} for the $C_{3,4}$ and $C_{7,8}$ operators.
}
\label{comboD7cd}
\end{figure}

Figures \ref{comboS6ab}-\ref{comboD7cd} compare the limits that can be derived for our effective operators
from the collider and indirect searches employed in this work. The most obvious feature of these plots is
the disparity in collider and indirect reaches for operators $C_{1-4}$, a result of the velocity suppression
of the bilinear $\bar{\chi}\chi$. Aside from this the collider and indirect reaches are seen to be comparable,
and highly complementary over the range of $m_{\chi}$ considered here.

\subsection{Heavy Boson Fiducial Efficiency and Limits}

\begin{figure}
\includegraphics[width=0.6\linewidth]{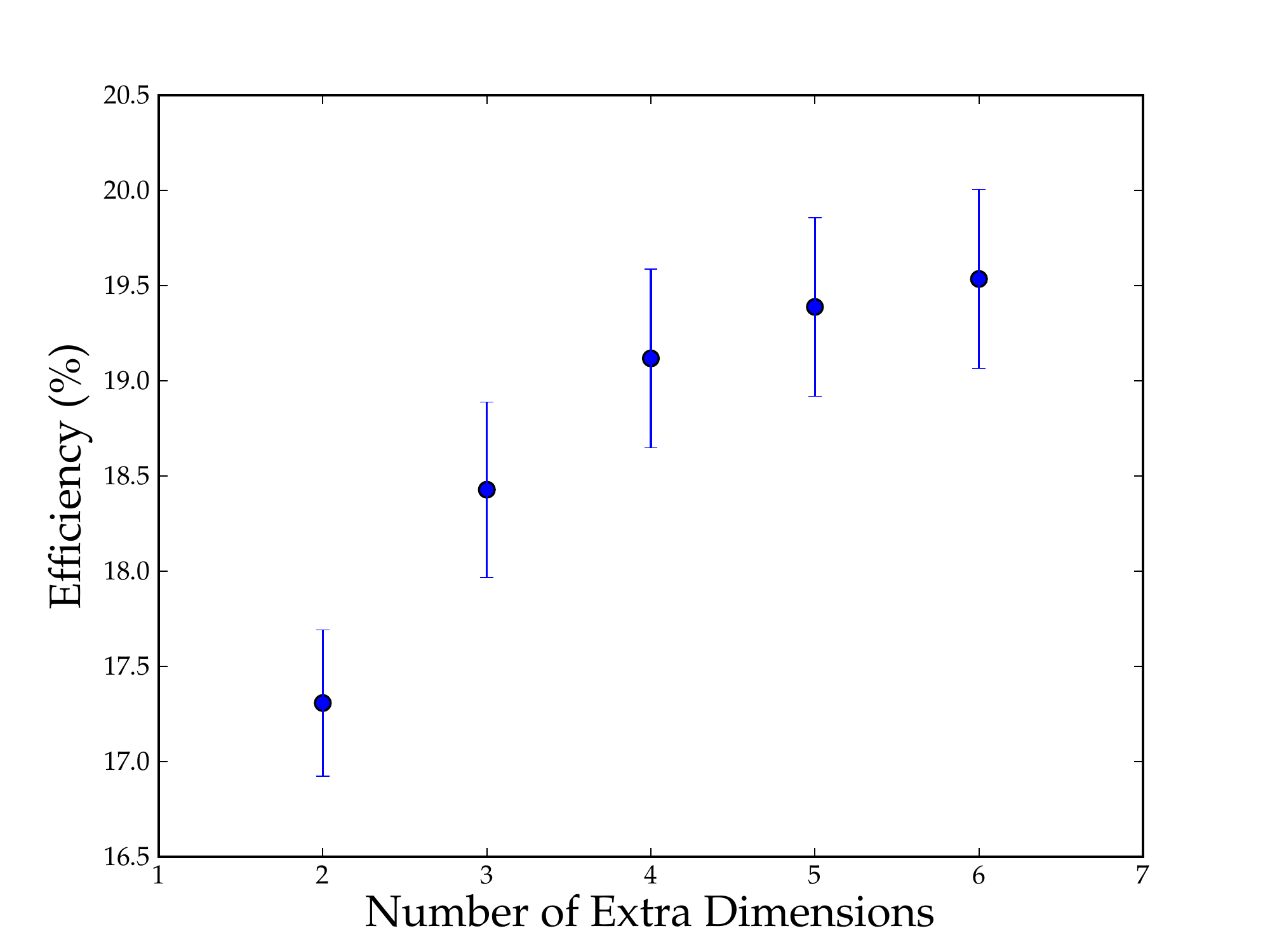}
\includegraphics[width=0.6\linewidth]{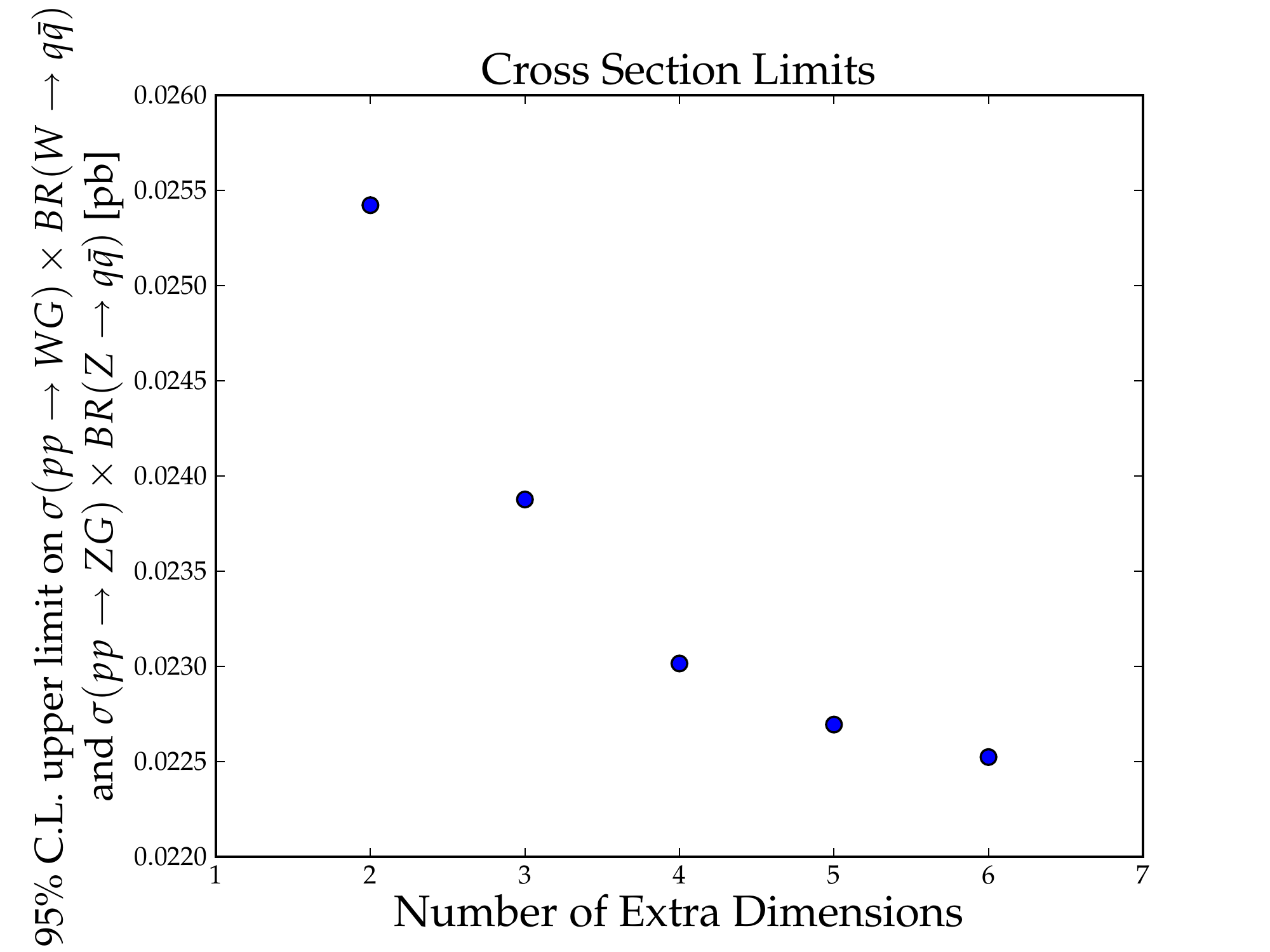}
\caption{ For  ADD~\cite{ArkaniHamed:1998rs} models predicting $WG$ or $ZG$ production, fiducial selection efficiency (top) and cross-section limits (bottom).}
\label{fig:grav}
\end{figure}

The fiducial efficiency  for $W\rightarrow GW$ and $Z\rightarrow GZ$ as a function of number of extra dimensions $\delta$ for the ADD model can be found in Fig.~\ref{fig:grav}(a). We find no significant difference for various $M_D$ values, so quote a single number for each $\delta$. 

As above, the efficiencies allow the derivation of limits on the cross section, see Fig.~\ref{fig:grav}(b).  These limits can be converted into limits on $M_D$ according to this relationship $\sigma \sim 1/M_D^{\delta+2}$, see Table~\ref{tab:MDlim}.   The current limits are also listed~\cite{CMScol}.

\begin{table}[tbh]
\caption{Limits on $M_D$ at 95\% CL calculated from limits on cross section in Fig.~\ref{fig:grav}(b) for each number of extra dimensions, along with the current limits on $M_D$~\cite{CMScol}.}
\center
\begin{tabular}{c|c|c}
\hline\hline
Number of Extra  & Our Limits   & Current  \\
 Dimensions & $M_D$ (TeV/$c^2$)  & Limits (TeV/$c^2$) \\\hline
2       & 1.84 & 5.67\\
3   	& 1.85 & 4.29\\
4       & 1.89 & 3.71\\
5       & 1.92 & 3.31\\
6       & 1.96 & 3.12\\
\hline  \hline
\end{tabular}
\label{tab:MDlim}
\end{table}

The most stringent current limits on $M_D$ in the ADD scenario come from CMS mono-jet searches~\cite{cmsjet}.  Our result is the first constraint on ADD derived in the mono-$W$ channel and it is competitive with other  hadron collider limits on $M_D$ in electroweak channels.  For example, CDF's mono-photon search for $\delta = 2$ places a limit on $M_D$ of 1.39 TeV \cite{pdg}.

\begin{figure}
\includegraphics[width=0.6\linewidth]{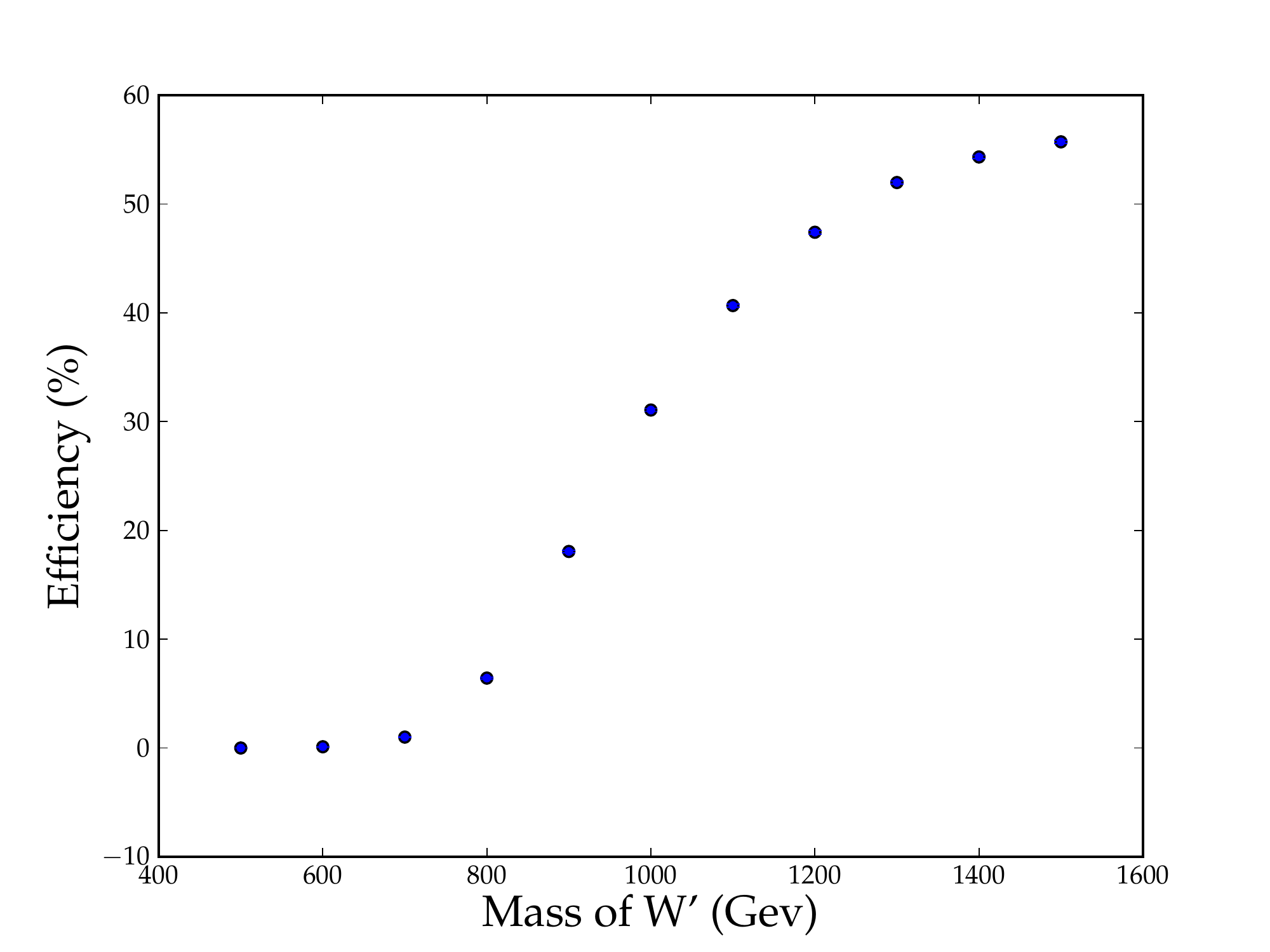}
\includegraphics[width=0.6\linewidth]{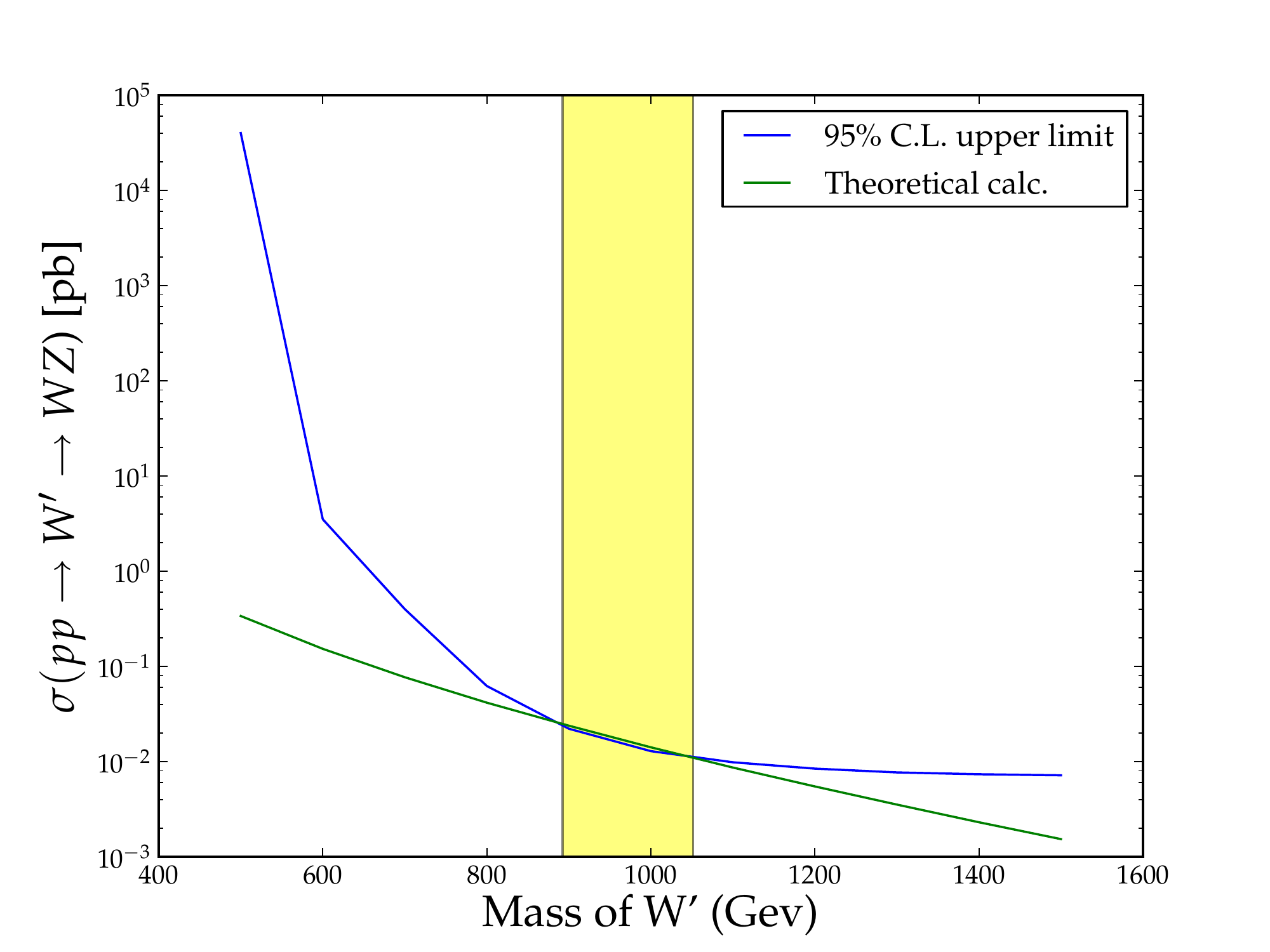}
\caption{ For models predicting $pp\rightarrow W'\rightarrow WZ$, fiducial selection efficiency (top) and cross-section limits for several values of $m_{W'}$ (bottom).}
\label{fig:Wp}
\end{figure}

In the case of the $W'$ boson, the selection efficiency is a strong function of the $W'$ boson mass, as the missing transverse energy is due to the $p_T$ of the $Z$ boson, which is approximately half of the $W'$ boson mass. The efficiency and cross section limits derived from it can be found in Fig.~\ref{fig:Wp}.

This excludes $W^{'}$ masses between 823 and 1055~GeV. This exclusion in the first for this scenario in the $W+\missET$ channel, and are a good complement to other search channels for fermiophobic scenarios.
Current limits on $W^{'}$ scenarios with large direct coupling to fermion are quite restrictive: 2.6 TeV for the $\ell \nu$ channel and 1.9 TeV for the $q \overline{q}$ \cite{pdg}.  However, limits in fermi-phobic scenarios are substantially more relaxed.  The most stringent limit is from CMS who searched in the multi-lepton channel $pp\rightarrow W'\rightarrow WZ$ where $W\rightarrow \ell \nu$ and $Z\rightarrow \ell \ell$;  in the sequential SM scenario the mass exclusion is 1.143 TeV~\cite{Chatrchyan:2012kk} assuming $g_{W^{'}WZ}/g_{WWZ}$= $(m_W/m_{W'})^2$. The ATLAS collaboration search in the same channel  yielded a limit of 760 GeV\cite{Aad:2012vs}. 

\subsection{Conclusions}

In this work we have derived constraints on dark matter interactions
with $W$ bosons in the context of a simply parameterized effective theory
framework. $W+\missET$ bounds derived by the ATLAS collaboration for
dark matter interactions with quarks were recast to find bounds on our model
for both scalar and fermionic dark matter scenarios, and compared to limits derived from indirect experiments.  We note that due to gauge invariance effective operators considered in this analysis which couples dark matter to pairs of W bosons must also predict non-zero coupling to of dark matter to other pairs of gauge bosons.  Thus we expect mono-W search constraints may be combined with other searches, for example mono-photons, to increase the power of the effective operator analysis. 

We have additionally pointed out that the $W+\missET$ have sensitivity in to heavy boson theories. We give results in terms of an ADD model of extra dimensions to produce limits on cross section and graviton coupling to $W$, placing lower limits on $M_D$ between 1.84 and 1.96 TeV. We analyze $W'$ production to give the first limits on this model in this final state, excluding $W'$ masses between 823 and 1055 GeV.  

\subsection{Acknowledgements}

We acknowledge useful conversations with Tim Tait.
DW and NZ are supported by grants from the Department of Energy
Office of Science. RC was supported in part by the National Science Foundation
under PHY-0970173 and PHY11-25915.

\clearpage
\appendix

\end{document}